\documentclass[11pt,twoside]{article}%
\usepackage{graphicx}%
\usepackage{amsmath}%
\usepackage{amsfonts}%
\usepackage{amssymb}

\begin{document}

\begin{center}
{\Large \textbf{Infrared extinction by aggregates of SiC particles: }}

{\Large \textbf{Comparison of different theoretical approaches }}

{\Large \ }

{\large Anja C. Andersen$^1$, Harald Mutschke$^2$, Thomas Posch$^3$ }

{\large \vspace{1.0076pc} }
{\small \textit{$^1$NORDITA, Blegdamsvej 17, DK-2100 Copenhagen, Denmark\\}}
{\small \textit{$^2$Astrophysikalisches Institut, Schillerg\"asschen 2-3, 
D-07745 Jena, Germany\\}}
{\small \textit{$^3$Institut f\"ur Astronomie, T\"urkenschanzstra{\ss}e 17, 
A-1180 Wien, Austria\\}}

{\small \textit{tel: (+45) 3532 5229, fax: (+45) 3538 9157 , e-mail: \vspace{1.0038pc}anja@nordita.dk }}
\end{center}

\section{{\protect\normalsize Abstract }}
Particle shape and aggregation have a strong influence on the spectral profiles 
of infrared phonon bands of solid dust grains. In this paper, we use a 
discrete dipole approximation (DDSCAT; {\normalsize \cite{draine+flatau04}}), 
a cluster-of-spheres code following the G\'erardy-Ausloos approach 
(MQAGGR; {\normalsize \cite{quinten93}}) and a T-matrix method 
(SCSMTM; {\normalsize \cite{mackowski96}}) for calculating IR extinction 
spectra of aggregates of spherical silicon carbide (SiC) particles. 
We compare the results obtained with the three different methods and 
discuss differences in the band profiles.

\section{{\protect\normalsize Introduction}}
Grain growth by aggregation is an important process in dense cosmic 
environments as well as in the earth's atmosphere. Besides influencing 
dynamic properties it also changes the absorption and scattering properties 
of the solid dust particles for electromagnetic radiation (e.g.\ {\normalsize 
\cite{stog95}}). This is especially true in spectral regions where resonant 
absorption occurs, such as the phonon bands in the infrared. It is 
quite well known that shape and aggregation effects actually determine 
the band profiles of such absorption and emission bands, which hinders 
e.g.\ the identification of particulate materials by their IR bands, 
but detailed investigations especially of the influence of grain 
aggregation are still lacking. We plan to set up a spectroscopic experiment 
(see Tamanai et al., this volume) for measuring aggregation effects on 
IR extinction by dust particles dispersed in air and, simultaneously, 
have started to use light scattering theory in order to predict 
numerically the band profiles for different aggregation states.  

\begin{figure}[b]
\center\includegraphics[width=12cm,angle=0]{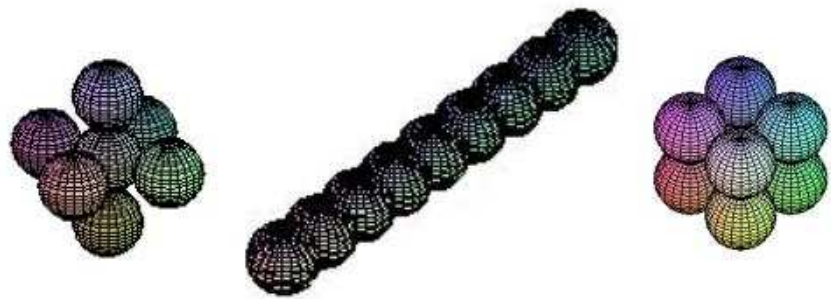}
\caption{The clusters frac7, lin9 and sc8.}
\label{clusteres}
\end{figure}

\section{{\protect\normalsize Structure of the clusters}} \label{structure_sect}
We consider three-dimensional clusters of identical touching spherical 
particles arranged in three different geometries: fractal, cubic, and linear. 
For a high precision of the calculations, we restrict the number of 
particles per cluster to less than 10. Therefore, we have selected only 
three geometries, namely a ``snowflake 1'st--order prefractal'' cluster 
(fractal dimension $D=\ln7/\ln3=1.77$, {\normalsize \cite{vicsek83}}), 
where one sphere is surrounded by six others along the positive and negative 
cartesian axes, a cluster of eight spheres arranged as a cube and 
a linear chain of nine spheres. All of the clusters (Fig.\,1) consist of 
spheres with radii $R=10$~nm and are embedded in vacuum (or air). 

 \begin{figure}[t]
 \centering\includegraphics[width=12cm,angle=0]{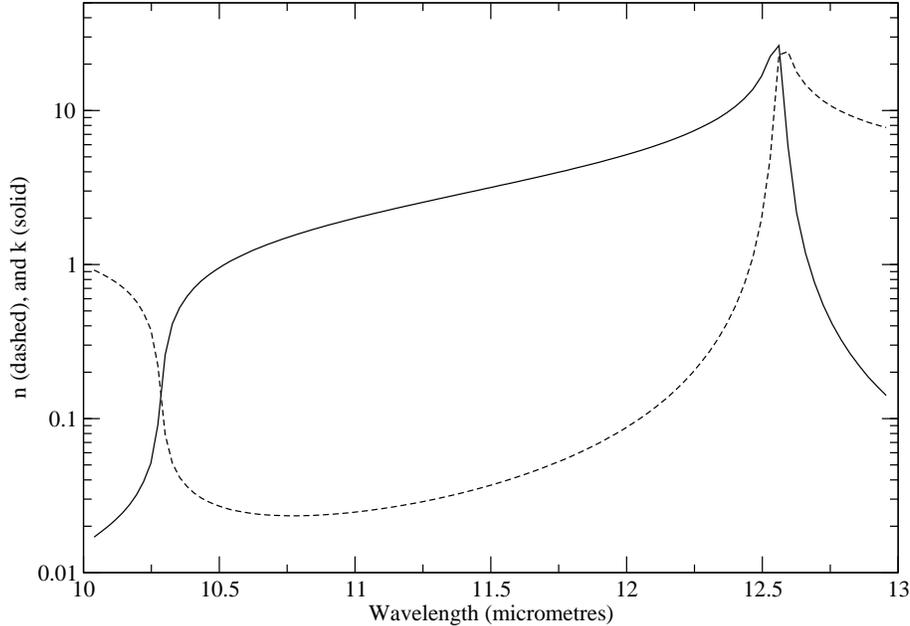}
 \caption{The refractive index for $\beta$-SiC with the damping constant
$\gamma$ = 2 cm$^{-1}$. The damping constant $\gamma$ (inverse proportional 
to the phonon life time) is an ``ad hoc'' introduced parameter, which 
in a perfect crystal reflects the anharmonicity of the potential curve. 
High-quality SiC is characterized by damping constants of 1--3 cm$^{-1}$.}
 \end{figure}

For the optical constants of the particle we have chosen the data of 
$\beta$-SiC in the wavelength range 10-13$\mu$m, calculated from 
a Lorentzian oscillator-type dielectric function describing the 
phonon resonance in this wavelength range (see {\normalsize \cite{mutschke99}}). 
On the one hand, this phonon resonance is of practical importance 
since it is observed as an emission band from dust particles in 
carbon star envelopes. On the other hand, depending on the resonance 
damping parameter, it represents a model material of a very high 
complex refractive index with $|m|>10$ and sharp surface resonances 
in the wavelength range between the LO and TO frequencies.

\section{{\protect\normalsize Computational approaches}}
\subsection{{\protect\normalsize The DDA method}}

The discrete dipole approximation (DDA) method is one of several 
discretisation methods (e.g.\ {\normalsize \cite{draine88}}, 
{\normalsize \cite{hage+greenberg}}) for solving scattering problems 
in the presence of a target with arbitrary geometry.
In this work we use the DDSCAT code version 6.1 {\normalsize \cite{draine+flatau04}}, 
which is very popular among astrophysicists.
In DDSCAT the considered grain/cluster is replaced by a cubic array
of point dipoles of certain polarizabilities {\normalsize \cite{goodman}}. 
The cubic array has numerical advantages because the conjugate 
gradient method can be efficiently applied to solve the matrix equation 
describing the dipole interactions. By specifying an appropriate 
grid resolution, calculations of the scattering and absorption of 
light by inhomogeneous media such as particle aggregates can be carried out 
to in principle whatever accuracy is required. 
For the sc8 cluster we used a grid of $36 \times 36 \times 36$ dipole, 
which provides 23\,752 dipoles in the cluster, while for the lin9 cluster 
a grid of $64 \times 64 \times 64$ dipole was used providing 
1676 dipoles in the cluster. The program performs orientational 
averaging of the clusters.

\begin{figure}[t]
\centering\includegraphics[width=12cm,angle=-90]{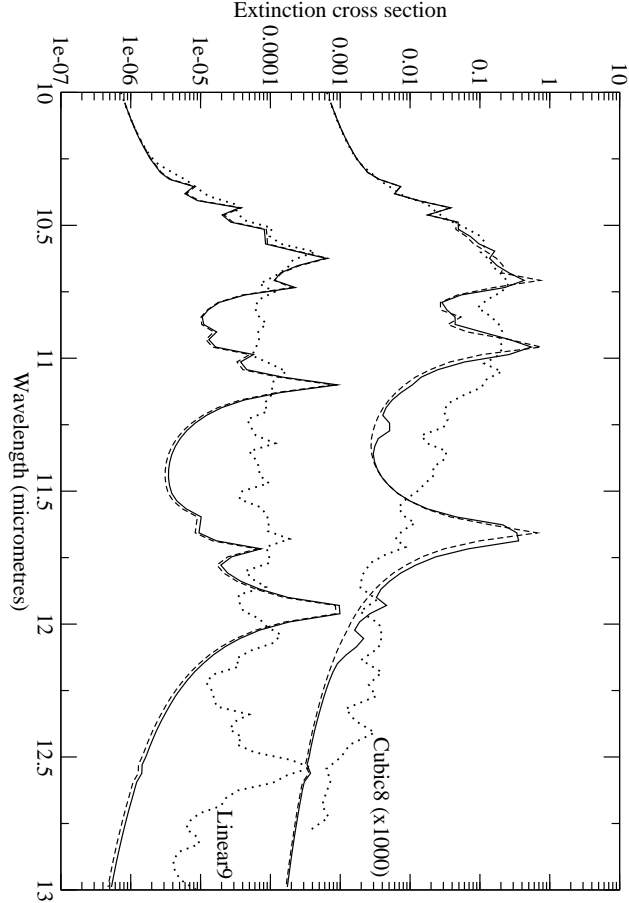}
\caption{Extinction profiles in the SiC phonon band for optical constants 
with $\gamma$ = 2 cm$^{-1}$
and two different cluster geometries, the linear chain of 9 touching spheres 
and the cubic cluster of 8 touching spheres. Dotted curves show the results 
obtained with DDSCAT, dashed curves with SCSMTM, and solid curves with MQAGGR.}
\end{figure}

\subsection{{\protect\normalsize The clusters-of-spheres method}}

The clusters-of-spheres calculations have been performed using (1) the 
program developed by M. Quinten (MQAGGR, commercially available) 
based on the theoretical approach by {\normalsize \cite{gerardy82}} and 
(2) using the T-matrix code by D.W. Mackowski (SCSMTM) calculating the 
random-orientation scattering matrix for an ensemble of spheres
{\normalsize \cite{mackowski96}}. Both programs aim at solving the scattering 
problem in an exact way by treating the superposition of incident and 
all scattered fields, developed into a series of vector spherical harmonics. 
Available computer power, however, forces to truncate the series at 
a certain maximum multipole order npol$_{max}$, which in both programs 
can be specified explicitly. Furthermore, both programs perform an 
orientational average of the cluster, the resolution of which was 
set to 15 degrees in MQAGGR and 10 degrees in SCSMTM for theta 
(the scattering angle). The variation in the azimuthal angle is not 
specified in SCSMTM. In MQAGGR it is varied betwen 0 and 360 degrees, 
again with a resolution of 15 degrees. 

\section{Results and discussion}

The extinction profiles shown in Fig.\,3 display the results obtained for 
two cluster geometries and three different codes. The profiles obtained with 
the two clusters-of-spheres codes at the same multipolar order (13 for the 
linear chain and 15 for the cubic cluster) give quasi-identical results in the 
case of the linear chain but differences for the cubic cluster. SCSMTM gave 
less resonances than MQAGGR, which is true for all lower maximum multipolar 
orders as well. The reason for that could possibly be different ways of 
orientational averaging. This is not completely understood yet. Generally 
speaking, both codes didn't converge up to the multipolar orders tried 
(up to 2h of CPU time per wavelength point). 

DDSCAT at the maximum resolution used (1676 dipoles in the linear cluster, 
23752 dipoles in the cubic one) gave much smoother profiles, i.e. with 
less distinct resonances. DDSCAT possesses limitations for large refractive 
indices ({\normalsize \cite{draine+goodman}}). On the other hand, the 
resonances produced by the clusters-of-spheres codes depend so much on 
the maximum multipolar order that it seems reasonable to assume that 
the (unreached) converged spectrum would show a more smooth profile 
as well. Results with lower dipole resolution tend to show sharp 
resonances as well.

\subsection*{{\protect\normalsize Acknowledgment}}
      We would like to thank B.T.\,Draine and P.J.\,Flatau for making their
      DDA code and D.W.\,Markowski for making his T-matix code available 
      as shareware.


\begin{thebibliography}{9}

\bibitem {draine+flatau04}B.T. Draine, P.J. Flatau, ``User guide for the discrete dipole approximation code DDSCAT 6.1'', http://arxiv.org/abs/astro-ph/030969 (2004).

\bibitem {quinten93}M. Quinten, U. Kreibig, ``Absorption and elastic scattering 
of light by particle aggregates'', Appl. Opt. \textbf{32}, 6173--6182 (1993).

\bibitem {mackowski96}D.W. Mackowski, M.I. Mishchenko, ``Calculation of the 
T matrix and scattering matrix for ensembles of spheres'', J. Opt. Soc. Amer. 
\textbf{A 13}, 266--2278 (1996).

\bibitem {stog95} R. Stognienko, Th. Henning, V. Ossenkopf, ``Optical
properties of coagulated particles'', Astron. Astrophys. \textbf{296}, 
797--809 (1995). 

\bibitem {vicsek83} T. Vicsek, ``Fractal models for diffusion controlled 
aggregation'', J. Phys. \textbf{A 16}, L647--L652 (1983). 

\bibitem {mutschke99} H. Mutschke, A.C. Andersen, D. Cl\'ement, Th. Henning, 
G. Peiter, ``Infrared properties of SiC particles'', Astron. Astrophys.
\textbf{345}, 187--202 (1999).

\bibitem {draine88} B.T. Draine, ``The discrete-dipole approximation and 
its application to interstellar graphite grains'', Astrophys. J. \textbf{333}, 
848--872 (1988).

\bibitem {hage+greenberg} J.I. Hage, J.M. Greenberg, ``A model for the 
optical properties of porous grains'', Astrophys. J. \textbf{361}, 
251--259 (1990). 

\bibitem {goodman} J.J. Goodman, B.T. Draine, P.J. Flatau, ``Application of FFT techniques to the discrete dipole approximation'', 
Opt.\ Lett.\ 16, 1198--1200 (1991).


\bibitem {gerardy82} J.M. G\'erardy, M. Ausloos, ``Absorption spectrum of 
clusters of spheres from the general solution of Maxwell's equations. 
II. Optical properties of aggregated metal spheres'', Phys. Rev. \textbf{B 25},
4204--4229 (1982)

\bibitem {draine+goodman} B.T. Draine, J.J. Goodman, ``Beyond Clausius-Mossotti - Wave propagation on a polarizable point lattice and the discrete dipole 
approximation'', Astrophys.\ J.\ \textbf{405}, 685--697 (1993).


\end{thebibliography}
\end{document}